\documentclass[conference]{IEEEtran}
\IEEEoverridecommandlockouts
\usepackage{cite}
\usepackage{amsmath,amssymb,amsfonts}
\usepackage{algorithmic}
\usepackage{graphicx}
\usepackage{textcomp}
\usepackage{xcolor}
\usepackage{nccmath}
\usepackage{lipsum}
\usepackage{subcaption}
\usepackage{multicol}
\usepackage{tempora}
\usepackage{url}
\usepackage[numbers]{natbib}
\usepackage{url}

\usepackage{breakurl}
\usepackage[breaklinks]{hyperref}
\usepackage{graphicx}
\newcommand{\RomanNumeralCaps}[1]{\MakeUppercase{\romannumeral #1}}
\begin{document}

\title{A Novel Approach to Alleviate Wealth Compounding in Proof-of-Stake Cryptocurrencies}

\author{\IEEEauthorblockN{Zahra Naderi}
\IEEEauthorblockA{\textit{School of Electrical and Computer Eng.} \\
\textit{College of Engineering} \\
\textit{University of Tehran} \\
zahranaderi976@ut.ac.ir}
\and
\IEEEauthorblockN{Seyed Pooya Shariatpanahi}
\IEEEauthorblockA{\textit{School of Electrical and Computer Eng.} \\
\textit{College of Engineering} \\
\textit{University of Tehran} \\
p.shariatpanahi@ut.ac.ir}
\and
\IEEEauthorblockN{Behnam Bahrak}
\IEEEauthorblockA{\textit{School of Electrical and Computer Eng.} \\
\textit{College of Engineering} \\
\textit{University of Tehran} \\
bahrak@ut.ac.ir}
}

\maketitle

\begin{abstract}
Due to its minimal energy requirement the PoS consensus protocol has become an attractive alternative to PoW in modern cryptocurrencies. In this protocol the chance of being selected as a block proposer in each round is proportional to the current stake of any node. Thus, nodes with higher stakes will achieve more block rewards, resulting in the so-called rich-getting-richer problem. In this paper, we introduce a new block reward mechanism called the FRD (Fair Reward Distribution) mechanism, in which for each block produced, in addition to a major reward given to the block proposer, a small reward is given to all other nodes. We prove that this reward mechanism makes the PoS protocol fairer in terms of concentration of wealth by developing on the Bagchi-Pal urn model.
\end{abstract}

\section{Introduction}
In the blockchain, there must be a mechanism to elect a node as a leader to propose the next block. Also, the elected leader must be given some stakes as a reward to incentivize nodes to participate in the blockchain procedure. PoW consensus mechanism \cite{nakamoto2008bitcoin,xiao2020survey,jakobsson1999proofs,gervais2016security}, which is used in Bitcoin cryptocurrency, is based on the processing power of the individuals. There is a puzzle that the first node to solve is selected as a leader, and the chance of being selected is proportional to the processing power of nodes. To alleviate the energy consumption problem of PoW \cite{zhang2020evaluation,Analysis}, in PoS \cite{xiao2020survey,saleh2021blockchain} the chance of becoming a leader is proportional to the percentage of shares in the stake pool. But this change leads to another challenge called the compounding of wealth phenomenon \cite{fanti2019compounding,mohman,Rammeloo,nair2021evaluation,zhang2020evaluation}. That is because the node with more stake has a higher chance of being selected as a leader. Upon being selected, it is given a reward, and its percentage of shares increases. So in the next time interval, its chance to be selected as a leader increases. Therefore, “the rich becomes richer, and the poor becomes poorer.”\par 
Concentration of wealth is a good indicator for observing fairness in PoS systems \cite{fanti2019compounding}. This article aims to reduce the variance of fractional stake distribution to improve fairness in the blockchain system by changing the block reward mechanism. In the proposed scheme, the final average stake of each node is proportional to its initial share. Moreover, the variance of final shares is significantly reduced compared to the conventional PoS system, approaching zero in the limit of the large-enough time interval. Our proposal is based on the Bagchi-Pal urn model introduced in \cite{bagchi1985asymptotic}.
\subsection{Related work}
Studying fairness of reward mechanism in cryptocurrencies has been the topic of many papers. Such papers have suggested different methods to measure fairness, and accordingly have reached different conclusions. For example, the authors in \cite{fanti2019compounding} introduce an equitability metric. Also, the authors in \cite{huang2021rich} propose two metrics called Expectational and Robust fairnesses to compare PoW and PoS, while \cite{wang2020incentive} has used the Gini coefficient to argue that PoS is fairer. Finally, the authors in \cite{karakostas2019cryptocurrency} have proposed egalitarian metrics which show that PoS is more equitable than PoW.\par
To improve fairness in PoS, some researchers introduced new consensus protocols aiming to improve fairness. Others tried to improve it by changing some aspects of the PoS protocol. In \cite{fanti2019compounding}, the authors first represented the equitability metrics and with minimizing it, found a new reward function named the Geometric reward, which varies with time. It reduces the variance of fractional stake distribution. \cite{wang2020incentive} introduced a new reward function called Bonus reward, in which the idea is adding salt (with exponential distribution with zero mean ) to the Geometric reward to make the reward curve smoother. In \cite{rocsu2021evolution} the authors argued that the time varying reward functions proposed in \cite{fanti2019compounding} is not stable along time and alternatively proposed to carefully adjust the initial shares to overcome the challenge. In \cite{huang2021rich} a solution based on withholding the rewards is proposed but no analytical proof is presented. Also, in \cite{brunjes2020reward}, another model is proposed for evaluating PoS fairness, through which it is argued that PoW is not fairer than PoS. Based on a game theoretical analysis of Reward-Sharing Schemes (RSS) fair stake pools are encouraged in PoS blockchains.\par
In our model, the proposed block reward is much like a constant reward independent of time, but the variance tends to zero over time which is an ideal fairness in the blockchain. However, in Geometric reward \cite{fanti2019compounding}, the variance of fractional stake distribution decreases but it does not reach zero. Additionally, in Geometric reward, the reward amount drops sharply after each time interval, which decreases the nodes' incentive to participate in the blockchain. In the Bonus reward \cite{wang2020incentive}, the sharpness of the Geometric reward decreases, but the distribution variance is still not as good as the FRD mechanism. The authers in \cite{rocsu2021evolution} introduce an easier way to improve fairness. However, for the variance to converge to zero, the amount of reward should converge to zero compared to the initial stake, which reduces the motivation of the nodes to participate in the blockchain. So in this model, there is a trade-off between incentive and fairness, but in the FRD mechanism, such a trade-off does not exist. Finally, the idea of reward withholding \cite{huang2021rich}, helps a lot to decrease variance, but it harms nodes’ motivation because nodes expect to get access to their rewards as soon as possible.\par
The structure of the paper is organized as follows. First, the system model have been explained in Section \RomanNumeralCaps{2}. Then, the new reward function with proof and simulation results have been provided to ensure that the FRD mechanism is fairer in Sections \RomanNumeralCaps{3} and \RomanNumeralCaps{4}. Finally, Section \RomanNumeralCaps{5} concludes the paper.
\section{System Model}
Assume that the model has $m$ node which are: $\mathcal{P}=\lbrace P_{1},...,P_{m}\rbrace$. In each round, each node's payout is added to the stake pool rather than being spent, and a new block is added to the chain. The entire stake owned by party $P_{i}$ in the proposed stake pool at time $n$ is denoted by $S_{P_{i}}(n)$, and the total stake in the proposer stake pool at time $n$ is denoted by $S(n)=\sum_{i=1}^{m} S_{P_{i}}(n)$. Moreover, $v_{P_{i}}(n)$ denotes the fractional stake of node $P_{i}$ at time $n$:
\begin{equation}
v_{P_{i}}(n)=\dfrac{S_{P_{i}}(n)}{S(n)}.
\end{equation}
The system selects a proposer node $G(n)\in P$ for each time slot $n$ as follows:
\begin{equation}
G(n) = \left\{
    \begin{array}{lll}
        P_{1} & \textrm{with probability}\; v_{P_{1}}(n)\\
        \vdots\\
        P_{m} &\textrm{with probability}\; v_{P_{m}}(n)\\
    \end{array}
\right..
\end{equation}
We define an $m\times m$ reward matrix $\textbf{R}=\left[ r_{i,j}\right]$ in which the $g$-th row indicates the rewards given to nodes if $G(n)=P_{g}$ as follows:
\begin{equation}
S_{P_{j}}(n+1)=S_{P_{j}}(n)+r_{g,j}\quad \textrm{for all}\; j\in \lbrace 1,...,m\rbrace.
\end{equation}
In contrast to previous works, we consider non-zero values for non-diagonal elements of $\textbf{R}$, which means upon selecting node $g$ as the proposer, all other nodes receive a reward as well. However, since the diagonal elements of $\textbf{R}$ are much larger than other elements, the proposer gets more reward than non-proposer nodes. The total reward in each time slot is fixed, as a result $\sum\limits_{i=1}^{m} r_{g,i} = K$ for all $g\in \lbrace 1,...,m\rbrace$, where $K$ is a constant. The conventional PoS Constant reward mechanism corresponds to $\textbf{R}=K\textbf{I}$ where $\textbf{I}$ is the identity matrix. In an ideal blockchain system in terms of fairness, each node’s expected fractional reward is approaching its initial
stake fraction: 
\begin{equation}
\mathbb{E}\left[ v_{P_{i}}(n)\right]=v_{P_{i}}(0),
\end{equation}
and the variance is approaching zero: 
\begin{equation}
Var\left[ v_{P_{i}}(n)\right]=0, 
\end{equation}
in the limit of the large-enough time interval.

\section{FRD mechanism Function}

\begin{figure*}[t] 
\centering
\begin{subfigure}{0.49\textwidth}
    \includegraphics[height=6.5cm,width=\textwidth]{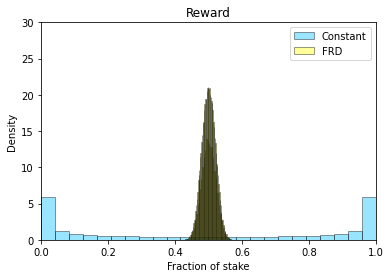}
    \caption{Initial stake ratio = 1/2}\label{subfig:p1}
\end{subfigure}
\begin{subfigure}{0.49\textwidth}
    \includegraphics[height=6.5cm,width=\textwidth]{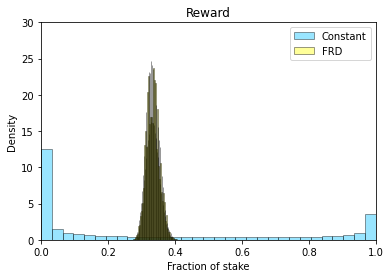}
    \caption{Initial stake ratio = 1/3}\label{subfig:p2}
\end{subfigure}      
\caption{Fractional stake distribution of the party A of Constant reward and FRD mechanism.}
\label{fig1}
\end{figure*}

\begin{table*}[t]
\caption{Probabilistic parameters of fractional stake distribution}
\begin{center}
\begin{tabular}{|p{0.28\textwidth}|p{0.15\textwidth}|p{0.15\textwidth}|p{0.15\textwidth}|p{0.15\textwidth}|}
\hline
 &\multicolumn{4}{|c|}{\textbf{Probabilistic parameters of reward function}} \\
\cline{2-5} 
\textbf{Initial values} & \textbf{Mean of Constant} & \textbf{Variance of Constant} & \textbf{Mean of FRD} & \textbf{Variance of FRD} \\
\hline
Four nodes, Initial stake ratio = $1/10$&$10^{-1}$&$5.99\times 10^{-2}$&$10^{-1}$&$1.54\times 10^{-4}$ \\
\hline
Ten nodes, Initial stake ratio = $1/10$&$9.95\times 10^{-2}$&$5.96\times 10^{-2}$&$10^{-1}$&$1.53\times 10^{-4}$ \\
\hline
Two nodes, Initial stake ratio = $1/2$&$4.99\times 10^{-1}$&$1.67\times 10^{-1}$&$5.00\times 10^{-1}$&$4.28\times 10^{-4}$\\
\hline
Two nodes, Initial stake ratio = $1/3$&$3.31\times 10^{-1}$&$1.47\times 10^{-1}$&$3.33\times 10^{-1}$&$3.83\times 10^{-4}$\\
\hline
\end{tabular}
\label{tab1}
\end{center}
\end{table*}

FRD is a PoS reward function based on the Bagchi-Pal urn model \cite{mahmoud2008polya} to alleviate the compounding of wealth phenomenon. First, the model is explained with an example of a blockchain system with two nodes as follows:
	$$
	\textbf{R}_{\textbf{C}} = \textbf{R}_{\textbf{Constant}} = 
	\begin{bmatrix} 
	200 & 0 \\
	0 & 200 \\
	\end{bmatrix},
	$$
	$$
	\textbf{R}_{\textbf{FRD}} =
	\begin{bmatrix} 
	150 & 50 \\
	50 & 150 \\
	\end{bmatrix}.
	$$
In the first case, when the party $A$ is elected as the proposer, it wins $200$ stakes, and nothing is given to the $B$ party. But in the FRD, when the party $A$ is elected as the proposer, it wins $150$ stakes, and $50$ stakes are given to the $B$ party. It should be noted that the above proposed reward matrix is for the case where the party $A$ has $1/2$ of the initial stake pool at the beginning (the total initial stakes are $200$). Fig.~\ref{fig1}-\subref{subfig:p1} shows the fractional stake distributions for $n = 10^{3}$ and approximately $10^{5}$ repetition.\par
If party $A$ has $1/3$ of the initial stake pool (the total initial stakes are $200$), the following reward functions are chosen and Fig.~\ref{fig1}-\subref{subfig:p2} shows the fractional stake distributions for $n = 10^{3}$ and approximately $10^{5}$ repetition.
	$$
	\textbf{R}_{\textbf{C}} =
	\begin{bmatrix} 
	200 & 0 \\
	0 & 200 \\
	\end{bmatrix},
	$$
	$$
	\textbf{R}_{\textbf{FRD}} =
	\begin{bmatrix} 
	133.33 & 66.67 \\
	33.33 & 166.67 \\
	\end{bmatrix}.
	$$
\noindent \par
In both simulations, the variance of the fractional stake of the FRD mechanism function is much less than the Constant reward, and the mean is equal to the node’s initial stake fraction in both of them. However, the fractional stake of the Constant reward follows Beta distribution according to \cite{fanti2019compounding}. So, the new model is a better solution to the problem of wealth accumulation.\par
Although the above example was for a system of two nodes, the same approach can be generalized as follows ($\alpha$ is an arbitrary coefficient):\\
$\textbf{R}_{\textbf{FRD}}=$
\begin{gather*}
	\alpha
	\begin{bmatrix} 
	S_{P_{1}}(0)+S(0) & S_{P_{2}}(0) &\cdots & S_{P_{m}}(0)\\
	S_{P_{1}}(0) & S_{P_{2}}(0)+S(0) &\cdots & S_{P_{m}}(0)\\
	\vdots & \vdots & \ddots & \vdots \\
	S_{P_{1}}(0) & S_{P_{2}}(0) &\cdots & S_{P_{m}}(0)+S(0)\\
	\end{bmatrix}
\end{gather*}
\begin{gather}\label{eq:a}
	= \alpha S(0)
	\begin{bmatrix} 
	v_{P_{1}}(0)+1 & v_{P_{2}}(0) &\cdots & v_{P_{m}}(0)\\
	v_{P_{1}}(0) & v_{P_{2}}(0)+1 &\cdots & v_{P_{m}}(0)\\
	\vdots & \vdots & \ddots & \vdots \\
	v_{P_{1}}(0) & v_{P_{2}}(0) &\cdots & v_{P_{m}}(0)+1\\
	\end{bmatrix}.
\end{gather}
The intuition behind above reward matrix is that the nodes with higher initial stakes deserve winning rewards even if they are not selected as the proposer. Also, the real proposer gets an additional reward proportional to the total initial stake. Theorem $1$ formally states the benefits of using the proposed reward matrix $\textbf{R}_{\textbf{Z}}$ in \eqref{eq:a}.\\
\textbf{Theorem 1.} The mean and variance of the number of node $i$’s stakes after $n$ steps are equal to:
\begin{equation}
\mathbb{E}\left[ S_{P_{i}}(n)\right] = v_{P_{i}}(0)2n + o(n),
\end{equation}
\begin{multline}
Var\left[ S_{P_{i}}(n)\right] = \alpha S(0)\left(2-v_{P_{i}}(0)\right) v_{P_{i}}(0)n\ln{n}\\
+ o(n\ln{n}).    
\end{multline}
In Corollary $1$ the main result of the paper in the limit of large steps is presented.\\
\textbf{Corollary 1.}
In the limit of long time, using the FRD mechanism function, the mean and variance of fractional stake distribution converge as follows:
\begin{gather}
\lim_{n \to +\infty} \mathbb{E}\left[ v_{P_{i}}(n)\right] = v_{P_{i}}(0),\\
\lim_{n \to +\infty} Var\left[ v_{P_{i}}(n)\right] = 0.
\end{gather}
It is an ideal result because the variance converges to zero, and each node's mean does not change in the long run.\\
\textbf{Proof}. Please refer to appendix.

\section{Simulation Results}
In this section, the simulation results are presented to verify the model's accuracy. Table ~\ref{tab1} summarize the numerical results.\par
\begin{figure*}[tp] 
\centering
\begin{subfigure}{0.49\textwidth}
    \includegraphics[height=6.5cm,width=\textwidth]{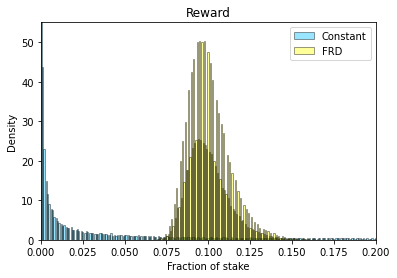}
    \caption{4 nodes}
\end{subfigure}
\begin{subfigure}{0.49\textwidth}
    \includegraphics[height=6.5cm,width=\textwidth]{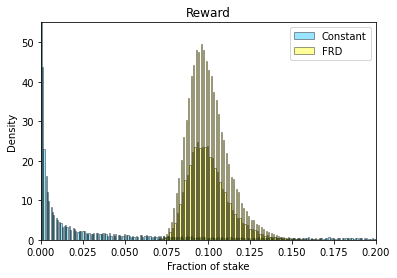}
    \caption{10 nodes}
\end{subfigure}      
\caption{Fractional stake distribution of Constant reward and FRD mechanism for multiple node.}
\label{fig3}

\vspace*{\floatsep}

\centering
\begin{subfigure}{0.49\textwidth}
    \includegraphics[width=\textwidth]{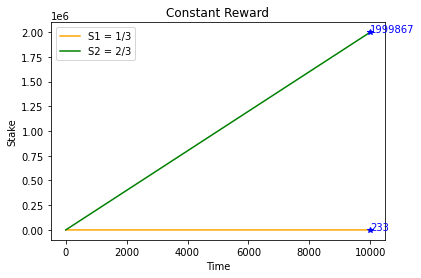}
    \caption{Constant reward}\label{subfig:a}
\end{subfigure}
\begin{subfigure}{0.49\textwidth}
    \includegraphics[width=\textwidth]{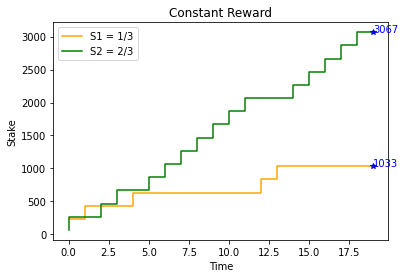}
    \caption{Constant reward (zoom in)}\label{subfig:b}
\end{subfigure}
\begin{subfigure}{0.49\textwidth}
    \includegraphics[width=\textwidth]{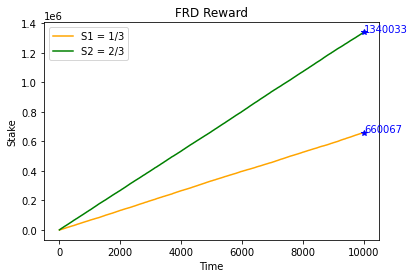}
    \caption{FRD mechanism}\label{subfig:c}
\end{subfigure}
\begin{subfigure}{0.49\textwidth}
    \includegraphics[width=\textwidth]{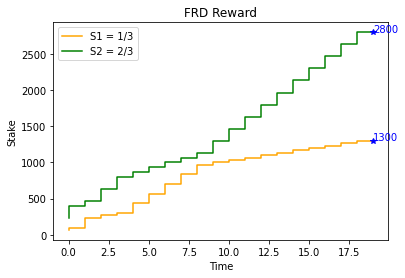}
    \caption{FRD mechanism (zoom in)}\label{subfig:d}
\end{subfigure}    
\caption{Total nodes stake over time}
\label{fig4}
\end{figure*}

Fig.~\ref{fig3} shows the result of the FRD mechanism for a system of multiple nodes, with the total initial stake $S(0)$ = $100$ and the total reward $K$ = $200$ in each time slot. The considered node was allocated $1/10$ of the total stakes at the beginning. The general trend of Fig.~\ref{fig3} is the same as Fig.~\ref{fig1} discussed earlier.\par
Fig.~\ref{fig4} shows the difference between Constant reward and FRD mechanism from another perspective for two nodes. The initially allocated stakes for the two nodes are $1/3$ and $2/3$. The figure shows the stakes of the nodes over time. In the Constant reward the node with the lower stake will get almost no further stakes over time while the richer node constantly accumulates wealth. However, in the FRD mechanism, the slope of node $2$ is twice of node $1$, as the initial stake ratio which is consistent with Theorem $1$. In order to get a better understanding of the dynamics of stake changes a zoomed version is also plotted in ~\ref{fig4}-\subref{subfig:b} and ~\ref{fig4}-\subref{subfig:d}.\par
\begin{figure*}[tp] 
\centering
\begin{subfigure}{0.49\textwidth}
    \includegraphics[trim=0 0 0 0.7cm, clip=true,height=5.8cm,width=\textwidth]{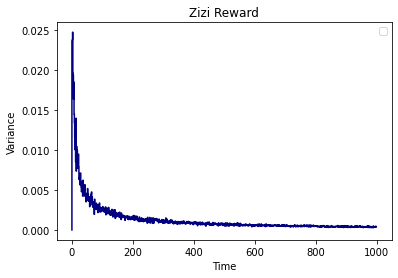}
    \caption{Variance}
\end{subfigure}
\begin{subfigure}{0.49\textwidth}
    \includegraphics[trim=0 0 0 0.7cm, clip=true,height=5.8cm,width=\textwidth]{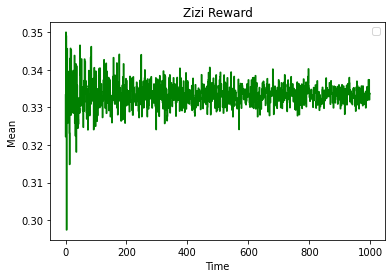}
    \caption{Mean}
\end{subfigure}      
\caption{Mean and Variance of FRD mechanism's Fractional stake distribution.}
\label{fig5}
\end{figure*}

In Fig.~\ref{fig5}, the mean and variance of a given node with initial stake share of $1/3$ is shown over time. In this simulation, two nodes exist, and the figure shows the one with $1/3$ of the total initial stake. The mean and variance were calculated for $100$ repetitions in each time slot. Fig.~\ref{fig5} verifies Theorem $1$ where variance converges to zero and mean to the initial stake ratio.

\section{Conclusion}
In this work, we study the effect of the compounding of wealth in the PoS protocol. The fractional stake distribution of Constant reward is like the Polya urn model which results in the well-known Beta distribution which is not satisfactory in terms of fairness. In contrast, we in present a new "FRD" mechanism to decrease the variance by developing on the Bagchi-Pal urn model. In this reward, the leader wins a constant amount of reward. Additionally, a little constant value is given to all nodes proportional to each node's initial stake ratio. We have shown mathematically and through simulations that this reward function drastically reduces the distribution variance and establishes equitability in the blockchain.

\bibliographystyle{unsrtnat}
\bibliography{ref}

\appendix
\textbf{Proof [Theorem 1].}
In this section,  the proof of Theorem is presented, but first for the working’s convenience, a variable change is applied to the reward function and theorem.\par
The Bagchi-Pal urn model \cite{mahmoud2008polya} which consists of an urn including white and black balls. It defines a reward matrix as below:
	$$
	\begin{bmatrix} 
	a & b \\
	c & d \\
	\end{bmatrix}.
	$$
In this model, if a white ball is selected, "$a$" white balls and "$b$" black balls are added to the urn. If a black ball is selected, "$c$" white balls and "$d$" black balls are added to the urn.\par
Next we consider a more general model suitable for our problem as follows. The reward matrix is defined as below ($w_{i}$ denotes $\alpha (S_{P_{i}}(0)+S(0))$ and $l_{i}$ denotes $\alpha S_{P_{i}}(0)$.):
	$$
	\begin{bmatrix} 
	w_{1} & l_{2} & l_{3} & \cdots& l_{m}\\
	l_{1} & w_{2} & l_{3} & \cdots& l_{m}\\
	l_{1} & l_{2} & w_{3} & \cdots& l_{m}\\
	\vdots & \vdots & \vdots & \ddots & \vdots \\
	l_{1} & l_{2} & l_{3} & \cdots& w_{m}\\
	\end{bmatrix},
	$$
where the sum of each row is equal to $K$.\par
Next we prove the mean of the number of node $i$’s stakes after $n$ steps is equal to:
\begin{equation}
\mathbb{E}\left[ S_{P_{i}}(n)\right] = \dfrac{l_{i}}{K-w_{i}+l_{i}}Kn + o(n).
\end{equation}
In order to calculate the variance, if $w_{i}-l_{i} < \dfrac{1}{2}K$ we have:
\begin{equation}
Var\left[ S_{P_{i}}(n)\right] = \dfrac{\left(K-w_{i}\right) l_{i}K\left(w_{i}-l_{i}\right) ^2}{\left(K-w_{i}+l_{i}\right) ^2\left(K-2\left(w_{i}-l_{i}\right) \right) }n + o(n),
\end{equation}
and if $w_{i}-l_{i} = \dfrac{1}{2}K$ we have:
\begin{equation}
Var\left[ S_{P_{i}}(n)\right] = \left(K-w_{i}\right) l_{i}n\ln{n} + o(n\ln{n}).
\end{equation}\par
The proof is along the same lines used in \cite{mahmoud2008polya,huang2021rich}, with some changes, and is provided here for completeness.\\
One can argue that:
\begin{gather*}
P(S_{P_{i}}(n+1)=S_{P_{i}}(n)+w_{i}|S_{P_{i}}(n)) = \dfrac{S_{P_{i}}(n)}{S(n)}=v_{P_{i}}(n),\\
P(S_{P_{i}}(n+1)=S_{P_{i}}(n)+l_{i}|S_{P_{i}}(n)) =1-v_{P_{i}}(n).
\end{gather*}
Its expectation is defined as follows:
\begin{multline*}
\mathbb{E}\left[ S_{P_{i}}(n+1)|S_{P_{i}}(n)\right]=\left(S_{P_{i}}(n)+w_{i}\right)  v_{P_{i}}(n)\\
+\left(S_{P_{i}}(n)+l_{i}\right) \left(1-v_{P_{i}}(n)\right) =\left(1+\dfrac{w_{i}-l_{i}}{S(n)}\right) S_{P_{i}}(n)+l_{i}.
\end{multline*}
As a result, the unconditional expectation is:
\begin{equation*}
\mathbb{E}\left[ S_{P_{i}}(n+1)\right]=\left(1+\dfrac{w_{i}-l_{i}}{S(n)}\right) \mathbb{E}\left[ S_{P_{i}}(n)\right]+l_{i},
\end{equation*}
for $n \geqslant 0$. We define the $Y_{n}$ random variable as follows:
\begin{equation*}
Y_{n} = S_{P_{i}}(n) - \dfrac{l_{i}}{K-w_{i}+l_{i}}S(n),
\end{equation*}
and insert in the previous equation:
\begin{equation*}
\mathbb{E}\left[ Y_{n+1}\right] = \left(1+\dfrac{w_{i}-l_{i}}{S(n)}\right) \mathbb{E}\left[ Y_{n}\right].
\end{equation*}
Solving the above recurrence we arrive at:
\begin{equation*}
\mathbb{E}\left[ Y_{n}\right] = \left(S_{P_{i}}(0) - \dfrac{l_{i}}{K-w_{i}+l_{i}}S(0)\right) \prod_{j=0}^{n-1}\left(1+\dfrac{w_{i}-l_{i}}{S(j)}\right).
\end{equation*}
According to \cite{milne2000calculus} as $n\rightarrow \infty$ we have:
\begin{equation*}
\prod_{j=0}^{n-1}\left(1+\dfrac{\left(w_{i}-l_{i}\right) }{S(j)}\right) = O\left(S(n)^{\dfrac{w_{i}-l_{i}}{K}}\right).
\end{equation*}
Since $w_{i}-l_{i} < K$, we conclude that:
\begin{equation*}
\mathbb{E}\left[ Y_{n}\right] = O\left(n^{\left(w_{i}-l_{i}\right) /K}\right) = o(n).
\end{equation*}
As a result, the unconditional expectation is:
\begin{multline*}
\mathbb{E}\left[ S_{P_{i}}(n)\right] = \dfrac{l_{i}}{K-w_{i}+l_{i}}S(n) + \mathbb{E}\left[ Y_{n}\right]\\
 = \dfrac{l_{i}}{K-w_{i}+l_{i}}Kn + o(n),
\end{multline*}
where the linear term is dominant.\\
Next, for calculating the variance one can argue that:
\begin{gather*}
P(S_{P_{i}}(n+1)^2=\left(S_{P_{i}}(n)+w_{i})^2|S_{P_{i}}(n)\right) = \dfrac{S_{P_{i}}(n)}{S(n)},\\
P(S_{P_{i}}(n+1) ^2=\left(S_{P_{i}}(n)+l_{i}) ^2|S_{P_{i}}(n)\right) =1- \dfrac{S_{P_{i}}(n)}{S(n)}.
\end{gather*}
Its expectation is defined as follows: 
\begin{multline*}
\mathbb{E}\left[ Y_{n+1}^2|Y_{n}\right]\\
=\left(Y_{n}+w_{i}-\dfrac{l_{i}K}{K-w_{i}+l_{i}}\right) ^2\left(\dfrac{l_{i}}{K-w_{i}+l_{i}}+\dfrac{Y_{n}}{S(n)}\right) \\
+\left(Y_{n}+l_{i}-\dfrac{l_{i}K}{K-w_{i}+l_{i}}\right) ^2\left(\dfrac{K-w_{i}}{K-w_{i}+l_{i}}-\dfrac{Y_{n}}{S(n)}\right) .
\end{multline*}
Therefore, the unconditional expectation is:
\begin{multline}\label{eq:b}
\mathbb{E}\left[ Y_{n+1}^2\right] = \left(1+\dfrac{2\left(w_{i}-l_{i}\right) }{S(n)}\right) \mathbb{E}\left[ Y_{n}^2\right]\\
+\dfrac{\left(K-w_{i}-l_{i}\right) \left(w_{i}-l_{i}\right) ^2}{\left(K-w_{i}+l_{i}\right) S(n)}\mathbb{E}\left[ Y_{n}\right] + \dfrac{\left(K-w_{i}\right) l_{i}\left(w_{i}-l_{i}\right) ^2}{\left(K-w_{i}+l_{i}\right) ^2}.
\end{multline}
First suppose $w_{i}-l_{i}<\dfrac{1}{2}K$, $w_{i}-l_{i}\neq 0$. Then we arrive at:
\begin{equation*}
\mathbb{E}\left[ Y_{n+1}^2\right] - \left(1+\dfrac{2\left(w_{i}-l_{i}\right) }{S(n)}\right) \mathbb{E}\left[ Y_{n}^2\right]=0,
\end{equation*} 
which has a solution:
\begin{equation*}
Q=\prod_{j=0}^{n-1}\left(1+\dfrac{2\left(w_{i}-l_{i}\right) }{S(j)}\right) ,
\end{equation*} 
and the particular solution is:
\begin{multline*}
R=\dfrac{\left(K-w_{i}\right) l_{i}\left(w_{i}-l_{i}\right) ^2}{\left(K-w_{i}+l_{i}\right) ^2\left(K-2\left(w_{i}-l_{i}\right) \right) }S(n)\\
- \left(\dfrac{K-w_{i}-l_{i}}{K-w_{i}+l_{i}}\right) \left(w_{i}-l_{i}\right) \mathbb{E}\left[ Y_{n}\right],
\end{multline*} 
so the complete solution can be written in the form:
\begin{equation*}
\mathbb{E}\left[ Y_{n+1}^2\right] = C.Q+R,
\end{equation*} 
where $C$ should satisfy:
\begin{equation*}
\mathbb{E}\left[ Y_{0}^2\right] = \left(S_{P_{i}}(0)-\dfrac{l_{i}S(n)}{K-w_{i}+l_{i}}\right) ^2.
\end{equation*}
The variance of $S_{P_{i}}(n)$ is: 
\begin{multline*}
Var\left[ S_{P_{i}}(n)\right] = \mathbb{E}\left[ S_{P_{i}}(n)^2\right]-\left(\mathbb{E}\left[ S_{P_{i}}(n)\right]\right) ^2\\
= \mathbb{E}\left[ Y_{n}^2\right]-\left(\mathbb{E}\left[ Y_{n}\right]\right) ^2.  
\end{multline*}
According to \cite{milne2000calculus} as $n\rightarrow \infty$ we have:
\begin{equation*}
\prod_{j=0}^{n-1}\left(1+\dfrac{2\left(w_{i}-l_{i}\right) }{S(j)}\right) = O\left(S(n)^{\dfrac{2\left(w_{i}-l_{i}\right) }{K}}\right).
\end{equation*}
With assumptions $w_{i}-l_{i} < \dfrac{1}{2}K$ and $S(n)=Kn + S(0)$, it results that:
\begin{multline*}
Var\left[ S_{P_{i}}(n)\right] = \dfrac{\left(K-w_{i}\right) l_{i}\left(w_{i}-l_{i}\right) ^2}{\left(K-w_{i}+l_{i}\right) ^2\left(K-2\left(w_{i}-l_{i}\right) \right) }S(n)\\
 + o(S(n))= \dfrac{\left(K-w_{i}\right) l_{i}K\left(w_{i}-l_{i}\right) ^2}{\left(K-w_{i}+l_{i}\right) ^2\left(K-2\left(w_{i}-l_{i}\right) \right) }n + o(n).
\end{multline*}
Next assume $w_{i}-l_{i} = \dfrac{1}{2}K$. Then equation \eqref{eq:b} is simplified to:
\begin{multline*}
\mathbb{E}\left[ Y_{n+1}^2\right] = \dfrac{S(n+1)}{S(n)}\mathbb{E}\left[ Y_{n}^2\right]+\dfrac{\left(K-w_{i}\right) ^2-l_{i}^2}{S(n)}\mathbb{E}\left[ Y_{n}\right]\\
+\left(K-w_{i}\right) l_{i}.
\end{multline*}
This arrives us at:
\begin{equation*}
\mathbb{E}\left[ Y_{n+1}^2\right] - \dfrac{S(n+1)}{S(n)}\mathbb{E}\left[ Y_{n}^2\right] = 0,
\end{equation*}
which has a solution $Q=S(n)$. With the substitution $R = S(n)g(n)+\left(w_{i}+l_{i}-K\right) \mathbb{E}\left[ Y_{n}\right], n\geq0$ we will arrive at the following recurrence:
\begin{equation*}
S(n+1)g(n+1)=S(n+1)g(n)+\left(K-w_{i}\right) l_{i},
\end{equation*}
where it is assumed: 
\begin{equation*}
g(0)=\dfrac{y_{0}^2+\left(K-w_{i}-l_{i}\right) y_{0}}{S(0)},
\end{equation*}
so that:
\begin{equation*}
g(n)=g(0) + \left(K-w_{i}\right) l_{i}\sum_{j=1}^{n}\dfrac{1}{S(j)}\qquad n\geq 1.
\end{equation*}
Thus for $n\geq 1$ and the constant $C$ we have:
\begin{multline*}
\mathbb{E}\left[ Y_{n}^2\right] = CS(n) + \left(w_{i}+l_{i}-K\right) \mathbb{E}\left[ Y_{n}\right]+S(n)g(0)\\
+S(n)\left(K-w_{i}\right) l_{i}\sum_{j=1}^{n}\dfrac{1}{S(j)}.
\end{multline*}
As $n\rightarrow \infty$:
\begin{equation*}
\sum_{j=1}^{n}\dfrac{1}{S(j)}\thicksim \dfrac{1}{K}\ln{S(n)}.
\end{equation*}
As a result, it is concluded that:: 
\begin{equation*}
\mathbb{E}\left[ Y_{n}^2\right] \thicksim \dfrac{\left(K-w_{i}\right) l_{i}}{K}S(n)\ln{S(n)},
\end{equation*}
which gives:
\begin{equation*}
Var\left[ S_{P_{i}}(n)\right] \thicksim \dfrac{\left(K-w_{i}\right) l_{i}}{K}S(n)\ln{S(n)}.
\end{equation*}\\
Thus, in the FRD mechanism matrix, if $w_{i}-l_{i} \leqslant \dfrac{1}{2}K$ then:
\begin{gather}
\lim_{n \to +\infty} \mathbb{E}\left[ v_{P_{i}}(n)\right] = \dfrac{l_{i}}{K-w_{i}+l_{i}},\\
\lim_{n \to +\infty} Var\left[ v_{P_{i}}(n)\right] = 0.
\end{gather}
\textbf{Proof [Corollary 1].}
The number of node $i$'s stakes must be divided by the total number of stakes to get the mean and variance of the fractional stake distribution of the node $i$ as following:
\begin{multline*}
\mathbb{E}\left[  \dfrac{S_{P_{i}}(n)}{Kn+S_{P_{i}}(0)}\right] = \dfrac{\mathbb{E}\left[ S_{P_{i}}(n)\right]}{Kn+S_{P_{i}}(0)} \\
= \dfrac{l_{i}}{K-w_{i}+l_{i}}\dfrac{Kn}{Kn+S_{P_{i}}(0)} + \dfrac{o(n)}{Kn+S_{P_{i}}(0)},
\end{multline*}
the linear term is dominant.
\begin{equation*}
\lim_{n \to +\infty} \mathbb{E}\left[  \dfrac{S_{P_{i}}(n)}{Kn+S_{P_{i}}(0)}\right] = \dfrac{l_{i}}{K-w_{i}+l_{i}}.
\end{equation*}
If $w_{i}-l_{i} < \dfrac{1}{2}K$:
\begin{gather*}
Var\left[  \dfrac{S_{P_{i}}(n)}{Kn+S_{P_{i}}(0)}\right] =  \dfrac{Var\left[ S_{P_{i}}(n)\right]}{\left(Kn+S_{P_{i}}(0)\right) ^2},\\
\lim_{n \to +\infty} Var\left[  \dfrac{S_{P_{i}}(n)}{Kn+S_{P_{i}}(0)}\right] = \lim_{n \to +\infty} \dfrac{n}{n^2}= 0,
\end{gather*}
and if $w_{i}-l_{i} = \dfrac{1}{2}K$:
\begin{gather*}
\lim_{n \to +\infty} Var\left[  \dfrac{S_{P_{i}}(n)}{Kn+S_{P_{i}}(0)}\right] = \lim_{n \to +\infty} \dfrac{n\ln{n}}{n^2} = 0.
\end{gather*}

\end{document}